\documentclass{article}
\usepackage{amssymb}

\usepackage{graphicx}
\usepackage{amsmath}

\input{tcilatex}

\begin{document}

\title{On dynamics of multi-phase elastic-plastic media}
\author{V. S. Borisov\thanks{%
The Pearlstone Center for Aeronautical Engineering Studies, Department of
Mechanical Engineering, Ben-Gurion University of the Negev, Beer-Sheva,
Israel. E-mail: viatslav@bgu.ac.il}}
\maketitle

\begin{abstract}
The paper is concerned with dynamics of multi-phase media consisting of a
solid permeable material and a compressible Newtonian fluid. Governing
macroscopic equations are derived starting from the space-averaged
microscopic mass and momentum balances. The Reynolds stress models (i.e.,
momentum dispersive fluxes) are discussed, and a suitable model is
developed. In the case of granular media the solid constituent is considered
as an elastic-plastic matrix, and the yield condition is approximated by
Coulomb friction law. It is revealed that the classical principle of maximum
plastic work is not, in general, valid for granular media, and an
appropriate variational principle is developed. This novel principle
coincides with the maximum plastic work principle for the case of
cohesionless granular media.
\end{abstract}

\section{Introduction}

Multi-phase mixtures play a vital part in many natural phenomena and
branches of engineering (e.g., \cite{Bear and Bachmat 1990}, \cite{Chan and
Lien 2005}, \cite{Jagering et el. 2001}, \cite{Kolev 2005a}, \cite{Kolev
2005b}, \cite{Nedderman 1992}, \cite{Nikolaevskij 1990}, \cite{Nikolaevskiy
1996}, \cite{Pan and Horne 2001}), and hence the development of multi-phase
dynamics is of great scientific and industrial importance. We restrict our
consideration to isotropic (e.g., \cite{Bear and Bachmat 1990}, \cite
{Nikolaevskii 1959}, \cite{Poreh 1965}) permeable (granular, porous, etc'.)
media consisting of a solid matrix and a compressible Newtonian fluid. Such
media have received the most study (see, e.g., \cite{Alam and Luding 2003}, 
\cite{Alam et al. 2005}, \cite{Bear and Bachmat 1990}, \cite{Didwania and
Boer 1999}, \cite{Nikolaevskij 1990}, \cite{Nikolaevskiy 1996}, \cite{Pan
and Horne 2001}, \cite{Prat et al. 2002}, \cite{Skjetne and Auriault 1999}),
nevertheless a number of important problems still remain to be solved. This
paper is mainly concerned with the two important problems, namely, modeling
of dispersive flux of momentum \cite{Bear and Bachmat 1990} as well as
development of a variational principle for plastic deformations \cite{Han
and Reddy 1999} of fluid-saturated granular media.

The so-called macroscopic balance equations are mainly developed by a method
of averaging (see, e.g., \cite{Bear and Bachmat 1990}, \cite{Dagan 1989}, 
\cite{Nikolaevskij 1990}, \cite{Nikolaevskiy 1996}) of micro-equations. The
method of averaging over elementary volume of a multi-phase medium
containing the full ensemble of realizations was originally suggested by
Nikolaevskiy et al. (1970) (see the references in \cite{Nikolaevskij 1990}, 
\cite{Nikolaevskiy 1996}). This approach, in contrast to the purely
phenomenological one (e.g., \cite{Jagering et el. 2001}, \cite{Nikolaevskiy
1996}) gives a possibility to evaluate theoretically the type of
constitutive laws and sometimes the values of rheological parameters \cite
{Nikolaevskij 1990}. However, because of non-linearity, the averaging
procedure leads to the macro-equations that are similar to Reynolds
equations (e.g., \cite{Anderson et al. 1984}, \cite{Monin and Yaglom 1971})
for turbulent flows, and hence the problem of Reynolds stress (i.e.,
dispersive flux of momentum \cite{Bear and Bachmat 1990}) modeling is coming
into play. In the fluid phase the pseudo-Brownian velocity pulsations exist
even under small Reynolds number as a result of chaotic micro-structure of
multi-phase media, and these pulsations are mainly changing in space \cite
{Nikolaevskiy 1996}. For such a motion the dispersivity tensor \cite{Poreh
1965} in the developed models of convective diffusion in, e.g., porous media 
\cite[sec. 7.3]{Nikolaevskij 1990} depends on the average velocity, whereas,
because of Galilei-Newton principle of relativity \cite[V. 2, p. 105]{Sedov
1971}, the turbulence model parameters do not depend on the velocity as such
(see, e.g., \cite{Monin and Yaglom 1971}, \cite{Anderson et al. 1984}).
Hence, the models of turbulent diffusion, as they exist, are not, in
general, applicable to multi-phase media. In the case of granular media the
Reynolds stresses (i.e., momentum dispersive fluxes) are often believed as
negligible \cite{Nikolaevskij 1990}. However, in developing models for
correct simulation of processes accompanying, e.g., fluid flow in vicinities
of gas wells having high production, underground explosion works, oil and
water wells rehabilitation and stimulation by shock technologies (steam
injection, aggressive pressure pulsing, etc.'), meteorological flows over
urban and vegetative canopies, and so on, the momentum dispersive fluxes
must be taken into consideration (e.g., \cite{Bear and Bachmat 1990}, \cite
{Chan and Lien 2005}, \cite{Katul et al. 2004}, \cite{Nikolaevskiy 1996}).

A dispersive flux model of an extensive quantity was suggested in \cite{Bear
and Bachmat 1990} for a microscopically laminar flow regime. The model is
essentially based on the modified rule \cite[Eq. 2.3.48]{Bear and Bachmat
1990} for volume averaging of a spatial derivative. Since the modified
averaging rule is also used in the development of the total viscous
resistance expression as well as the macroscopic momentum balance equation
for a fluid phase \cite[sec. 2.6]{Bear and Bachmat 1990}, we consider the
applicability of this rule to a viscous fluid flow through a permeable
medium.

The authors \cite{Bear and Bachmat 1990} attempted to develop the modified
rule for a quantity, G, that attains no maximum or minimum value within the
void space of a representative elementary volume (REV). On this basis it is
assumed \cite[p. 125]{Bear and Bachmat 1990} that the quantity G is a
harmonic function on the microscopic level. This demand is sufficient but
not necessary condition for the validity of the maximum principle, and hence
it could be too restrictive. To demonstrate it we assume, for the sake of
simplicity, that the solid matrix is immobile and the entire void space is
occupied by a single Newtonian fluid with the density $\rho =const$ and the
dynamic viscosity $\mu =const$. Eliminating the body force of gravity by
subtraction from the true pressure $p$ of the hydrostatic pressure, the
Navier-Stokes and continuity equations (e.g., \cite{Anderson et al. 1984}, 
\cite{Monin and Yaglom 1971}, \cite{Schlichting and Gersten 2000}) can be
written in the following non-dimensional form: 
\begin{equation}
S_{h}\frac{\partial \mathbf{V}}{\partial t}+\left( \mathbf{V}\cdot \nabla
\right) \mathbf{V=}-E_{u}\nabla P+\frac{1}{Re}\nabla ^{2}\mathbf{V},
\label{I-2}
\end{equation}
\begin{equation}
\nabla \cdot \mathbf{V}=0,\mathbf{\quad }S_{h}\equiv \frac{l_{\ast }}{%
V_{\ast }t_{\ast }},\mathbf{\ }E_{u}\equiv \frac{P_{\ast }}{\rho V_{\ast
}^{2}},\ Re\equiv \frac{V_{\ast }l_{\ast }}{\nu },  \label{I-4}
\end{equation}
where $\mathbf{V}$, $P$ are the non-dimensional velocity and pressure,
respectively; $\nu $ denotes the kinematic viscosity; $S_{h}$, $E_{u}$, $Re$
are, respectively, Struhal, Euler, and Reynolds numbers; the reference
quantities are denoted by an asterisk. Multiplying (\ref{I-2}) by $\nabla $
we obtain, in view of the first equation in (\ref{I-4}), that 
\begin{equation}
\nabla ^{2}P=-\frac{1}{E_{u}}\nabla \cdot \left[ \left( \mathbf{V}\cdot
\nabla \right) \mathbf{V}\right] .  \label{I-6}
\end{equation}
Thus, the basic equality, i.e., the Laplace equation for the pressure 
\cite[Eq. 2.6.6]{Bear and Bachmat 1990}, can be approximately valid if the
right-hand side in (\ref{I-6}) is negligible, i.e., in general, if $E_{u}\gg
1$. Hence, we conclude that the derivation of the viscous resistance
expression as well as the macroscopic momentum balance equation \cite[sec.
2.6]{Bear and Bachmat 1990}, relying on the modified averaging rule, can be,
in general, valid for a creeping flow only.

The applicability of the modified averaging rule for the development of the
momentum dispersive flux model is based on the assumption that the linear
momentum density ($\rho \mathbf{V}$) is a harmonic function on the
microscopic level. In view of (\ref{I-2}) we obtain that $\rho \mathbf{V}$
will be an approximate solution of the Laplace equation if the following
inequalities will be valid simultaneously: $S_{h}Re\ll 1$, $Re\ll 1$, $%
E_{u}Re\ll 1$. In general, it is possible for a very specific flow. As an
example, let us consider the Newtonian fluid flow (see the governing
equations in (\ref{I-2}) and (\ref{I-4})) through a porous medium as made up
of a bundle of parallel tubes whose radii are assumed to be uniform in size.
In the case of a steady-state laminar flow regime we have (in every tube),
in fact, Hagen-Poiseuille flow \cite[p. 117]{Schlichting and Gersten 2000}.
In polar coordinates the velocity ($V$) distribution over a cross-section of
a tube can be written in the form 
\begin{equation}
V=\frac{\zeta R^{2}}{4\mu }\left( 1-\frac{r^{2}}{R^{2}}\right) ,\quad 0\leq
r\leq R,\quad \zeta =\frac{P_{0}-P_{L}}{L},  \label{I-10}
\end{equation}
where $R$ denotes the radius of the tube; $L$ denotes the full length of the
tube; $\zeta $ denotes the pressure drop; $P_{0}$, $P_{L}$ are the pressures
at the bases of the tube. Obviously, the linear momentum density ($\rho V$)
in the case of Hagen-Poiseuille flow is not a harmonic function. To
elucidate the conditions wherein $\rho V$ is an approximate solution of the
Laplace equation we assume that the reference velocity is equal to the
maximum value of the velocity in Hagen-Poiseuille flow (i.e., $V_{\ast
}=0.25\zeta R^{2}/\mu $), $l_{\ast }=R$, and $P_{\ast }=P_{0}-P_{L}$. In
such a case we obtain that $E_{u}Re=4L/R$, and hence $\rho V$ will be an
approximate solution of the Laplace equation if $R\gg L$. The last
inequality is to say that the influence of the boundary (non-slip)
conditions on the flow regime must be negligible. This is by no means the
case of porous media. Thus, we conclude that the applicability of the
modified averaging rule to the linear momentum density ($\rho \mathbf{V}$)
is, in general, questionable.

Using the modified averaging rule as well as a number of additional
assumptions and approximations, Bear and Bachmat \cite{Bear and Bachmat 1990}
found that the dispersive flux of an extensive quantity is proportional to
the gradient of the mean density of the extensive quantity. Hence, the
momentum dispersive flux is proportional to the gradient of the mean
momentum density of mass. The mean momentum density in its turn may be
decomposed into two fluxes: a macroscopic advective flux and a dispersive
flux of mass. The latter flux, in view of the model by Bear and Bachmat \cite
{Bear and Bachmat 1990}, is proportional to the gradient of the mean density
of mass. Thus, following \cite{Bear and Bachmat 1990}, we obtain an
interesting result that the macroscopic momentum balance equation is a
third-order partial differential equation. However, in view of the foregoing
analysis, the basis for this result is not convincing. Two more points need
to be made. The momentum dispersive flux (see, e.g., left-hand side in 
\cite[Eq. 2.6.52]{Bear and Bachmat 1990}), which is a symmetrical tensor, is
approximated by the tensor (see the right-hand side in \cite[Eq. 2.6.52]
{Bear and Bachmat 1990}), which is not, in general, symmetric. Furthermore,
the kinetic energy of dispersion (analog to the kinetic energy of
turbulence, see \cite[p. 221]{Anderson et al. 1984}) is not taken into
consideration in the momentum dispersive flux model in \cite{Bear and
Bachmat 1990}.

Let us note, however, that the approach employed by Bear and Bachmat \cite
{Bear and Bachmat 1990} deserves more attention. This approach and the
conclusions, as it is underlined by the authors, are equally valid for any
extensive quantity. It is a reflection of the plausible assumption that the
mechanisms of heat, mass, and momentum transfer in multi-phase media are
identical. In the semi-empirical theory of turbulence the similar assumption
is called as Reynolds analogy (see, e.g., \cite{Anderson et al. 1984}, \cite
{Monin and Yaglom 1971}, \cite{Schlichting and Gersten 2000}). Specifically,
Anderson et al. \cite{Anderson et al. 1984} pointed out that the ratio of
the diffusivities for the turbulent transport of heat and momentum
(turbulent Prandtl number) is a well-behaved function across the flow, and
the Prandtl number varies between about 0.6 at the outer edge of the
boundary layer to about 1.5 near the wall. Assuming that this assumption is
valid, we can conclude \cite{Bear and Bachmat 1990} that the coefficient of
mechanical dispersion (dispersivity tensor, see \cite{Poreh 1965}) has
analogous form for any conservative extensive quantity, be it heat, mass, or
linear momentum.

It should also be remarked that Bear and Bachmat \cite{Bear and Bachmat 1990}
suggested the totally irreversible model of the momentum dispersive flux for
the case of microscopically laminar flow regime provided the density is
constant. In contrast to \cite{Bear and Bachmat 1990}, a reversible model of
the Reynolds stress tensor (i.e., the momentum dispersive flux) is developed
by Nikolaevskiy \cite{Nikolaevskiy 1996}. It is assumed that the local
velocity ($\mathbf{V}$) is a stochastic function of the average velocity ($%
\overline{\mathbf{V}}$). In such a case the vector of mean velocity is
transformed randomly (see \cite{Nikolaevskii 1959}, \cite{Nikolaevskiy 1996}%
) into the velocity pulsations ($\mathbf{V}^{\prime }$ $\equiv $ $\mathbf{V-}%
\overline{\mathbf{V}}$). This stochastic transformation is represented, in
both the symbolic and indicial notations, as follows: 
\begin{equation}
\mathbf{V}^{\prime }=\mathbf{L}\cdot \overline{\mathbf{V}};\quad
V_{i}^{\prime }=L_{ij}\overline{V}_{j}\,,  \label{I-25}
\end{equation}
where the repeated indices, as usual, denote summation; the tensor $\mathbf{L%
}$ is determined by the structure of porous medium, $Re$, and a realization
parameter, corresponding to the random character of the medium. Then, the
equations for the Reynolds stresses are, in fact, written \cite[sec. 4.4.1]
{Nikolaevskiy 1996} in the following form 
\begin{equation}
\overline{\mathbf{V}^{\prime }\mathbf{V}^{\prime }}=\mathbf{T}:\left( 
\overline{\mathbf{V}}\,\overline{\mathbf{V}}\right) ;\quad \overline{%
V_{i}^{\prime }V_{j}^{\prime }}=T_{ijkl}\overline{V}_{k}\overline{V}%
_{l},\quad T_{ijkl}=\overline{L_{ik}L_{jl}}\,,  \label{I-30}
\end{equation}
To estimate the kinetic energy of dispersion, $0.5\overline{V_{i}^{\prime
}V_{i}^{\prime }}$, we assume, in view of the isotropy of the porous medium 
\cite{Nikolaevskii 1959}, that the second rank tensor with the components $%
\overline{L_{ik}L_{il}}$ will be isotropic. Then, by virtue of (\ref{I-30}),
we obtain 
\begin{equation}
\overline{\left( \mathbf{V}^{\prime }\right) ^{2}}\equiv \overline{%
V_{i}^{\prime }V_{i}^{\prime }}=tr\overline{V_{i}^{\prime }V_{j}^{\prime }}=%
\overline{L_{ik}L_{il}}\ \overline{V}_{k}\overline{V}_{l}=\omega \left( 
\overline{\mathbf{V}}\right) ^{2},\quad \omega =\frac{1}{3}\mathbf{L:L.}
\label{I-40}
\end{equation}

Let us assess the validity of the assumption (\ref{I-25}) and, hence, the
model (\ref{I-30}). Notice, if the direction of $\overline{\mathbf{V}}$ in (%
\ref{I-25}) will be changed to the opposite one (i.e., $\overline{\mathbf{V}}
$ $\rightarrow $ $-\overline{\mathbf{V}}$), then the only direction of $%
\mathbf{V}^{\prime }$ will be changed (i.e., $\mathbf{V}^{\prime }$ $%
\rightarrow $ $-\mathbf{V}^{\prime }$). However, such a property is not, in
general, exhibited by the flow, e.g., in convergent and divergent channels.
The exact solution of the Navier-Stokes equations for such a flow was
originally found by Jeffery and Hamel (see the brief sketch in \cite[pp.
104-106]{Schlichting and Gersten 2000}). The velocity distribution for the
convergent and for the divergent channel differ significantly from each
other, and in the latter case vary greatly with Reynolds number (see, e.g., 
\cite{Frick 2003}, \cite{Schlichting and Gersten 2000}). Inasmuch as the
flow in such channels is essentially irreversible, we conclude that (\ref
{I-25}) can, in general, be approximately valid under low Reynolds numbers
only.

Frick \cite{Frick 2003} points out that a symmetric divergent Jeffery-Hamel
flow exists only if the Reynolds number $Re<\widehat{Re}$ and the opening
angle $\alpha <\widehat{\alpha }$, where the values $\widehat{Re}$ and $%
\widehat{\alpha }$ meet the following condition: $\widehat{Re}$ $=$ $6\left(
\pi ^{2}\diagup \widehat{\alpha }-\widehat{\alpha }\right) $. Therefore,
choosing $\alpha \geq \pi $ we obtain at least one region of back-flow,
whichever $Re$ might be, and, resulting from it, separation. Separation of a
boundary layer, in reality, gives rise to vortices \cite[Sec. 2]{Schlichting
and Gersten 2000} resulting in turbulence. For instance, measurements
demonstrate \cite[Sec. 7.2.6]{Schlichting and Gersten 2000} that the flow of
a free jet can be laminar up until about $Re=30$, where $Re$ is referred to
the outlet velocity and the slit height. Hence, in a real granular medium,
the solid phase of which composed, in general, of irregular in size and
shape grains, the vortices (i.e., micro-vortices) can arise under low
Reynolds numbers as the result of steep rise in pressure at sharp edges,
fractures, sudden expansions, etc'. The higher $Re$, the more micro-vortices
arise within an REV. Nevertheless, the flow is still laminar. Let us note,
however, that the vortices are the main source of turbulence \cite
{Belotserkovsky and Ginevsky 1995}. The turbulent flow stems from the lose
of vortex stability and degradation of vortex structure on further rise in $%
Re$. If $Re$ is over a critical Reynolds number, then the vortex structure
breaks into turbulence within a part of the REV. The higher $Re$, the most
part of the flow will be turbulent. The above speculation is supported by
experimental results (see, e.g., \cite{Bear and Verruijt 1987}, \cite
{Bennethum and Giorgi 1997}, \cite{Lage et al. 2002} and references
therein), and hence it might be differentiated the following regimes of
flow: 1) Laminar regime, where the resistance to the flow is directly
proportional to the mean velocity (Darcy linear law). 2) Laminar regime,
where the resistance is nonlinear (Darcy-Hazen-Dupuit-Forchheimer law). 3)
Transition regime. 4) Turbulent flow. Thus, from the preceding, it appears
that the assumption (\ref{I-25}) and, hence, the model (\ref{I-30}) can be
approximately valid in the case of the first flow regime (laminar regime,
Darcy linear flow), otherwise the validity of (\ref{I-25}) is questionable.
The same conclusion is valid for the model developed by Bear and Bachmat 
\cite{Bear and Bachmat 1990}.

Recently, several turbulence models have been established for turbulent
flows (i.e., for the fourth regime) in \ permeable (granular, porous, etc'.)
media (see, e.g., \cite{Chan and Lien 2005}, \cite{Katul et al. 2004}, \cite
{de Lemos and Mesquita 2003}, \cite{Nakayama and Kuwahara 1999}, \cite
{Pedras and de Lemos 2003}, \cite{Vadasz and Olek 1999}, \cite{Zhou and
Pereira 2000}, and references therein). Let us note that the widely used
one- and two-equation turbulence models (e.g., \cite{Chan and Lien 2005}, 
\cite{Katul et al. 2004}, \cite{de Lemos and Mesquita 2003}, \cite{Nakayama
and Kuwahara 1999}, \cite{Pedras and de Lemos 2003}, \cite{Zhou and Pereira
2000}) are only valid in the fully turbulent regime, i.e., these models are
not appropriate for the near-wall region \cite[pp. 231-233]{Anderson et al.
1984}. Owing to this, the validity of these models for turbulent flows in
porous media is questionable. Thus, as already noted in \cite[p. 63]{Kolev
2005a}, modeling of the Reynolds stresses for multi-phase flows is still at
its infancy.

The present study (Sec. \ref{Momentum transport}) is devoted to the
development of a sufficiently simple mathematical model for simulating flows
in\ permeable media under all regimes. To avoid the specific problems of
dispersive flux modeling associated with inconstant phase densities (e.g., 
\cite{Anderson et al. 1984}, \cite{Bear and Bachmat 1990}), we will develop
a Reynolds form of the balance equations in mass-averaged variables \cite[p.
201]{Anderson et al. 1984}. Then we will develop a model of the dispersive
flux assuming that: (i) the Reynolds stresses are linearly dependent on the
mean velocity derivatives (see, e.g., \cite{Anderson et al. 1984}, \cite
{Bear and Bachmat 1990}, \cite{Kolev 2005a}, \cite{Monin and Yaglom 1971}, 
\cite{Nikolaevskiy 1996}), (ii) the Reynolds stress tensor will be isotropic
if the mean velocity is constant \cite{Kolev 2005a}, (iii) the Reynolds
analogy is valid for multi-phase media.

Considering the granular medium as an elastic-plastic one (e.g., \cite
{Jagering et el. 2001}, \cite{Jiang and Liu 2003}, \cite{Nedderman 1992}, 
\cite{Nikolaevskij 1990}, \cite{Nikolaevskiy 1996}), we are concerned with
validity of the classical principle of maximum plastic work, as applied to
granular media. It can be seen from (e.g., \cite[p. 58]{Han and Reddy 1999}, 
\cite[V. 4, ch. 12]{Sedov 1971}) that the associative flow rule (the
normality law \cite[p. 58]{Han and Reddy 1999}) is the necessary condition
for validity of the maximum plastic work principle. Nikolaevskiy \cite
{Nikolaevskij 1990}, \cite{Nikolaevskiy 1996} has revealed that in the case
of granular media the irreversible strain-rate must be determined by the
non-associative flow rule. Thus, we may conclude that the principle of
maximum plastic work \cite{Han and Reddy 1999} is not in general valid for
granular media. Importance of variational principles in physics, including
the maximum plastic work principle developed by von Mises, Taylor, and
Bishop and Hill, is well known (e.g., \cite{Gyarmati 1970}, \cite{Han and
Reddy 1999}, \cite{Sadovskii 1997}, \cite{Sadovskii 1997}). In particular,
Han and Reddy \cite{Han and Reddy 1999} wrote, ''The principle of maximum
plastic work is a vital constituent of the theory of plasticity.'' Moreover,
currently, numerical methods for solving the problems of elastic-plastic
deformations, including construction of monotone (e.g., \cite{Borisov and
Sorek 2004}, \cite{Borisov and Mond 2007}) difference schemes, are based on
non-classic formulation of variational principles, namely variational
inequalities (\cite{Han and Reddy 1999}, \cite{Panagiotopoulos 1985}, \cite
{Sadovskii 1997}). Hence, there is a need to develop a proper variational
principle for plastic deformation of granular media. With this in mind we
will assume the validity of the Coulomb yield condition (e.g., \cite
{Nedderman 1992}, \cite{Nikolaevskij 1990}, \cite{Nikolaevskiy 1996}) and
the non-associative flow rule (e.g., \cite{Nikolaevskij 1990}, \cite
{Nikolaevskiy 1996}). Then the desired variational principle will be
rigorously deduced on the basis of irreversible thermodynamics (e.g., \cite
{Gyarmati 1970}, \cite{Nikolaevskij 1990}), as applied to granular media.
Let us note that in such a development one would like to use a theoretical
premise instead of the Coulomb friction law that is nothing more than an
empirical relationship \cite{Nedderman 1992}. Currently another approach,
free of Coulomb condition, is suggested by Jiang and Liu \cite{Jiang and Liu
2003}. The novel elastic theory \cite{Jiang and Liu 2003} accounts for
mechanical yield by a feature of non-linear elasticity only. However, from
physical point of view, such an approach is not well founded, as it is not
obvious that solid friction at inter-particle contacts can be totally
accounted for by non-linear elasticity. Furthermore, Jiang and Liu \cite
{Jiang and Liu 2003} developed their theory for the case of cohesionless
granular media only. Hence, the use of this theory \cite{Jiang and Liu 2003}%
, as it exists, for development of the variational principle is out of
question.

\section{Momentum transport\label{Momentum transport}}

We start our investigation with the volume averaged balance equations \cite
{Bear and Bachmat 1990}. In the absence of phase transition, the macroscopic
mass and, respectively, momentum balance equations for a fluid phase can be
written in the form 
\begin{equation}
\frac{\partial }{\partial t}\left( \phi \overline{\rho _{f}}\right) =-\nabla
\cdot \left( \phi \overline{\rho _{f}\mathbf{V}_{f}}\right) ,  \label{S10}
\end{equation}
\begin{equation*}
\frac{\partial }{\partial t}\left( \phi \overline{\rho _{f}\mathbf{V}_{f}}%
\right) =-\nabla \cdot \phi \left( \overline{\rho _{f}\mathbf{V}_{f}\mathbf{V%
}_{f}}-\overline{\mathbf{\sigma }_{f}}\right)
\end{equation*}
\begin{equation}
+\frac{\phi }{U_{f}}\int\limits_{S_{fs}}\mathbf{\sigma }_{f}\cdot \mathbf{n}%
ds+\phi \overline{\rho _{f}\mathbf{g}_{f}},  \label{S20}
\end{equation}
where $\rho _{f}$ denotes the fluid density, $\phi $ denotes the void
fraction, $\mathbf{V}_{f}$ denotes the velocity of the fluid, $\mathbf{%
\sigma }_{f}$ ($\equiv -p_{f}\mathbf{I}+\mathbf{\tau }_{f}$) denotes the
fluid stress tensor, $p_{f}$ denotes the pressure, $\mathbf{I}$ denotes the
identity tensor, $\mathbf{\tau }_{f}$ denotes the viscous stress tensor, $%
\mathbf{g}_{f}$ denotes the body force vector, $U_{f}$ denotes the volume
occupied by the fluid phase within an REV, $S_{fs}$ denotes the surface of
the fluid-solid interface, $\mathbf{n}$ denotes the outward unit vector to $%
U_{f}$ on $S_{fs}$. Let $U_{s}$ denote the volume occupied by the solid
phase within the REV, and let $e_{f}$ and $e_{s}$ denote any variables
referred to the fluid and solid phases, respectively. Hereinafter $\overline{%
e_{f}}$ and $\overline{e_{s}}$ denote the volume averages of $e_{f}$ and $%
e_{s}$ over, respectively, $U_{f}$ and $U_{s}$. Let $\widetilde{e_{f}}$ ($%
\equiv \overline{\rho _{f}e_{f}}\diagup \overline{\rho _{f}}$) and $%
\widetilde{e_{s}}$ ($\equiv \overline{\rho _{s}e_{s}}\diagup \overline{\rho
_{s}}$) denote the mass-averaged variables. Following Nikolaevskiy \cite
{Nikolaevskij 1990} we write the phase interaction force ($\mathbf{F}$) in
the form 
\begin{equation}
\mathbf{F}\equiv \frac{\phi }{U_{f}}\int\limits_{S_{fs}}\mathbf{\sigma }%
_{f}\cdot \mathbf{n}ds-\overline{p}\nabla \phi =-\mathbf{R}\cdot \widetilde{%
\mathbf{U}},\ \mathbf{U}\equiv \mathbf{V}_{f}-\widetilde{\mathbf{V}_{s}},
\label{S40}
\end{equation}
where $\mathbf{R}$ (=$\mu \psi \phi \mathbf{K}^{-1}$) denotes the symmetric
resistance tensor \cite{Nikolaevskij 1990}, $\mathbf{K}$ denotes the tensor
of absolute permeability, $\psi =\psi \left( Re\right) $, $Re$ ($\equiv
\left| \widetilde{\mathbf{U}}\right| l_{\ast }\diagup \overline{\nu }$)
denotes the local Reynolds number, 
$\nu =\mu \diagup \rho _{f}$ denotes the kinematic viscosity, $l_{\ast }$
denotes the length parameter characterizing the void space.

Using the mass-averaged variables and by virtue of (\ref{S40})\ we obtain
from (\ref{S10}), (\ref{S20}) the following balance equations: 
\begin{equation}
\frac{\partial }{\partial t}\left( \phi \overline{\rho _{f}}\right) =-\nabla
\cdot \left( \phi \overline{\rho _{f}}\widetilde{\mathbf{V}_{f}}\right) ,
\label{S50}
\end{equation}
\begin{equation*}
\phi \overline{\rho _{f}}\frac{d}{dt}\widetilde{\mathbf{V}_{f}}=-\phi \nabla 
\overline{p_{f}}-\mathbf{R}\cdot \left( \widetilde{\mathbf{V}_{f}}-%
\widetilde{\mathbf{V}_{s}}\right)
\end{equation*}
\begin{equation}
+\phi \overline{\rho _{f}}\widetilde{\mathbf{g}_{f}}+\nabla \cdot \phi 
\overline{\mathbf{\tau }_{f}}-\nabla \cdot \phi \overline{\rho _{f}}%
\widetilde{\mathbf{V}_{f}^{\prime \prime } \mathbf{V}_{f}^{\prime \prime }},
\label{S60}
\end{equation}
where $e_{f}^{\prime \prime }\equiv $ $e_{f}-\widetilde{e_{f}}$ denotes the
fluctuations of a variable $e_{f}$. We will also use $e_{f}^{\prime }=$ $%
e_{f}-\overline{e_{f}}$. Since the fluid phase is a Newtonian liquid, we can
write (e.g., \cite{Bear and Bachmat 1990}, \cite{Sedov 1971}) that 
\begin{equation}
\mathbf{\tau }_{f}=2\mu \mathbf{S}-\kappa \nabla \cdot \mathbf{V}_{f}\mathbf{%
I} ,\quad \mathbf{S}=\frac{1}{2}\left[ \nabla \mathbf{V}_{f}+\left( \nabla 
\mathbf{V}_{f}\right) ^{\ast }\right] ,  \label{S65}
\end{equation}
where $\kappa $ ($=\frac{2}{3}\mu -\zeta $) denotes the coefficient of bulk
viscosity, $\zeta $ denotes the second coefficient of viscosity, $\left( %
\mbox{\hspace{2.0mm}}\right) ^{\ast }$ denotes a conjugate tensor.
Estimating the mean value of the viscous tensor ($\overline{\mathbf{\tau }%
_{f}}$) we note \cite{Anderson et al. 1984} that in practice the viscous
terms involving $\overline{\mathbf{V}_{f}^{\prime \prime }}\equiv $ $-%
\overline{\rho _{f}^{\prime }\mathbf{V}_{f}^{\prime }}\diagup \overline{\rho
_{f}}$ can be neglected, and hence we obtain 
\begin{equation}
\overline{\mathbf{\tau }_{f}}\approx 2\overline{\mu }\widetilde{\mathbf{S}}- 
\overline{\kappa }\nabla \cdot \widetilde{\mathbf{V}_{f}}\mathbf{I},\quad 
\widetilde{\mathbf{S}}=\frac{1}{2}\left[ \nabla \widetilde{\mathbf{V}_{f}}
+\left( \nabla \widetilde{\mathbf{V}_{f}}\right) ^{\ast }\right] .
\label{S70}
\end{equation}
Notice, the first equality in (\ref{S70}) may be considered as strict
equation introducing new characteristics $\overline{\mu }$ and $\overline{%
\kappa }$\ instead of the mean value of conventional coefficients of
viscosity. In such a case we obtain that $\overline{\nu }=\overline{\mu }%
\diagup \overline{\rho _{f}}$.

The dispersion in the case of isotropic granular media is, in general,
non-isotropic \cite{Bear and Bachmat 1990}, which is to say that at every
point of the flow there must be defined a symmetric tensor ($\mathbf{D}_{f}$%
) of second rank (e.g., \cite{Bear and Bachmat 1990}, \cite{Monin and Yaglom
1971}, \cite{Nikolaevskij 1990}, \cite{Poreh 1965}) such that the dispersive
flux of the momentum will be dependent on $\mathbf{D}_{f}\cdot \nabla 
\widetilde{\mathbf{V}_{f}}$. Since the Reynolds stresses form a symmetric
tensor, it is natural to consider this tensor as a linear function of the
following symmetric one 
\begin{equation}
\mathbf{\Phi }_{f}\equiv \frac{1}{2}\left[ \mathbf{D}_{f}\cdot \nabla 
\widetilde{\mathbf{V}_{f}}+\left( \mathbf{D}_{f}\cdot \nabla \widetilde{%
\mathbf{V}_{f}}\right) ^{\ast }\right] .  \label{S80}
\end{equation}
We will assume that because of chaotic micro-structure of the isotropic
granular medium, the Reynolds stress tensor will be isotropic if $\mathbf{%
\Phi }_{f}=0$. Thus, we may write that 
\begin{equation}
-\overline{\rho _{f}}\widetilde{\mathbf{V}_{f}^{\prime \prime }\mathbf{V}%
_{f}^{\prime \prime }}=\overline{\rho _{f}}\left( \mathbf{\Phi }_{f}+d%
\mathbf{I}\right) ,  \label{S75}
\end{equation}
where $d$\ is a scalar parameter. In view of the last equality in (\ref{S40}%
) we obtain that $\mathbf{V}_{f}^{\prime \prime }$ $=$ $\mathbf{U}^{\prime
\prime }$. Then, assuming that 
\begin{equation}
\widetilde{\left| \mathbf{U}\right| ^{2}}\approx \left( 1+\omega _{f}\right)
\left| \widetilde{\mathbf{U}}\right| ^{2},  \label{S75-10}
\end{equation}
where $\omega _{f}$ ($\geq 0$) is a new non-dimensional parameter, we can
write 
\begin{equation*}
\widetilde{\left( \mathbf{V}_{f}^{\prime \prime }\right) ^{2}}=\widetilde{%
\mathbf{U\cdot \mathbf{U}}}\mathbf{-}2\widetilde{\mathbf{U\cdot }\widetilde{%
\mathbf{U}}}\mathbf{+}\widetilde{\mathbf{U}}\cdot \widetilde{\mathbf{U}}=%
\widetilde{\left( \mathbf{U}\right) ^{2}}-\left( \widetilde{\mathbf{U}}%
\right) ^{2}\approx
\end{equation*}
\begin{equation}
\omega _{f}\left( \widetilde{\mathbf{U}}\right) ^{2}=\omega _{f}\left( 
\widetilde{\mathbf{V}_{f}}-\widetilde{\mathbf{V}_{s}}\right) ^{2}.
\label{S76}
\end{equation}
It is often assumed (see, e.g., \cite{Bear and Bachmat 1990}, \cite
{Nikolaevskij 1990}) that $\widetilde{\left( \mathbf{U}\right) ^{2}}$ $%
\approx $ $\left( \widetilde{\mathbf{U}}\right) ^{2}$ (or $\overline{\left( 
\mathbf{U}\right) ^{2}}$ $\approx $ $\left( \overline{\mathbf{U}}\right)
^{2} $), and hence $\omega _{f}\approx 0$. Notice, such an assumption is not
valid for laminar regimes, however, $\omega _{f}$ can be close to zero for
the case of turbulent regime under sufficiently high Reynolds number (see
Sec. \ref{Appendix}).

Equating linear invariants of the tensors in (\ref{S75}) we obtain, in view
of (\ref{S76}), the following estimation of the Reynolds stress tensor 
\begin{equation}
-\overline{\rho _{f}}\widetilde{\mathbf{V}_{f}^{\prime \prime }\mathbf{V}%
_{f}^{\prime \prime }}=\overline{\rho _{f}}\mathbf{\Pi }_{f},  \label{S85}
\end{equation}
\begin{equation}
\mathbf{\Pi }_{f}=\mathbf{\Phi }_{f}-\frac{1}{3}\left( \omega _{f}\left| 
\widetilde{\mathbf{V}_{f}}-\widetilde{\mathbf{V}_{s}}\right| ^{2}+tr\mathbf{%
\Phi }_{f}\right) \mathbf{I}.  \label{S90}
\end{equation}

Using the Reynolds analogy (e.g., \cite{Anderson et al. 1984}, \cite{Monin
and Yaglom 1971}), as applied to multi-phase media, and following \cite
{Nikolaevskii 1959} and \cite{Poreh 1965} we find, for the case of isotropic
granular media, that the dispersivity tensor $\mathbf{D}_{f}$ can be
approximated as follows: 
\begin{equation}
D_{f;ij}=F_{1}\nu \delta _{ij}+F_{2}\frac{l_{\ast }^{2}}{\nu }\widetilde{U}%
_{i}\widetilde{U}_{j},  \label{S120}
\end{equation}
where 
\begin{equation}
F_{1}=\frac{a_{f}^{0}Re^{2}}{1+\alpha _{f}Re},\quad F_{2}=\frac{b_{f}}{%
1+\beta _{f}Re},\quad Re=\left| \widetilde{\mathbf{U}}\right| \frac{l_{\ast }%
}{\overline{\nu }},  \label{S130}
\end{equation}
$a_{f}^{0}$, $b_{f}$, $\alpha _{f}$, and $\beta _{f}$ denote the
dimensionless parameters describing the geometry of the void space, $%
\widetilde{U}_{i}$ $=$ $\widetilde{V_{f;i}}-\widetilde{V_{s;i}}$. We find,
by virtue of (\ref{S120})-(\ref{S130}), that

\begin{equation}
D_{f;ij}=\left( a_{f}l_{\ast }\left| \widetilde{\mathbf{U}}\right| \delta
_{ij}+b_{f}l_{\ast }\widetilde{U}_{i}\widetilde{U}_{j}\diagup \left| 
\widetilde{\mathbf{U}}\right| \right) \frac{Re}{1+\beta _{f}Re},
\label{S110}
\end{equation}
where $a_{f}=a_{f}^{0}\left( 1+\beta _{f}Re\right) \diagup \left( 1+\alpha
_{f}Re\right) $. The similar approximation of the dispersivity tensor is
suggested in \cite[p. 218]{Bear and Bachmat 1990}. Notice, the approximation
(\ref{S110}) of the dispersivity tensor $\mathbf{D}_{f}$ does not contradict
with Galilei-Newton principle of relativity \cite[V. 2, p. 105]{Sedov 1971},
since $\mathbf{D}_{f}$ depends on $\widetilde{\mathbf{U}}\equiv \widetilde{%
\mathbf{V}_{f}}-\widetilde{\mathbf{V}_{s}}$.

Similarly we obtain the macroscopic balance equations for the solid phase: 
\begin{equation}
\frac{\partial }{\partial t}\left[ \left( 1-\phi \right) \overline{\rho _{s}}%
\right] =-\nabla \cdot \left[ \left( 1-\phi \right) \overline{\rho _{s}}%
\widetilde{\mathbf{V}_{s}}\right] ,  \label{S150}
\end{equation}
\begin{equation}
\left( 1-\phi \right) \overline{\rho _{s}}\frac{d}{dt}\widetilde{\mathbf{V}%
_{s}}=\nabla \cdot \overline{\mathbf{\sigma }}^{f}-\left( 1-\phi \right)
\nabla \overline{p_{f}}+\left( 1-\phi \right) \overline{\rho _{s}}\widetilde{%
\mathbf{g}}_{s}+\mathbf{R}\cdot \left( \widetilde{\mathbf{V}_{f}}-\widetilde{%
\mathbf{V}_{s}}\right) ,  \label{S160}
\end{equation}
where $\overline{\mathbf{\sigma }}^{f}=$ $\left( 1-\phi \right) \left( 
\overline{\mathbf{\sigma }}_{s}+\overline{p_{f}}\mathbf{I}\right) $ is the
Terzaghi effective stress \cite{Nikolaevskij 1990}. Notice, following
Nikolaevskiy \cite{Nikolaevskij 1990}, the Reynolds stress tensor (i.e.,
momentum dispersive flux \cite{Bear and Bachmat 1990}), $\left( 1-\phi
\right) \overline{\rho _{s}}\widetilde{\mathbf{V}_{s}^{\prime \prime }%
\mathbf{V}_{s}^{\prime \prime }}$, is assumed as negligible for the solid
phase. In such a case we obtain the conventional mathematical model for
elastic-plastic deformations of granular media, i.e., the system of
hyperbolic equations. However, as it can be concluded, e.g., from \cite{Bear
and Bachmat 1990}, the dispersion of the momentum should, in general, be
taken into consideration if irreversible deformations take place. Then, in
view of the Reynolds analogy, the momentum dispersive flux for the solid
phase can be estimated by application of (\ref{S85}), (\ref{S90}), (\ref{S80}%
), and (\ref{S110}) with obvious modifications. In such a case we obtain the
system of partial differential equations of parabolic type as a mathematical
model of multi-phase dynamics.

\section{Maximum principle}

Hereinafter, the considered quantities, such as stress, density, etc.', will
be of the average ones only. Hence, the sign to indicate the fact of
averaging will be deleted.

To develop a variational principle for plastic deformation of granular media
we start with Coulomb friction law, as applied to a two-phase granular
medium. According to Terzaghi's principle \cite{Nikolaevskij 1990} the yield
condition is formulated to the effective stresses: 
\begin{equation}
C_{\sigma }\equiv \frac{2}{\sqrt{3}}\left| \sigma _{\tau }\right| +\alpha
\sigma ^{f}-Y=0,\quad \sigma ^{f}=\frac{1}{3}tr\mathbf{\sigma }^{f},
\label{S180}
\end{equation}
where $\sigma _{\tau }$ denotes the shear stress intensity in the solid
matrix, $\alpha $ ($=\alpha \left( \chi \right) >0$) denotes the internal
friction coefficient, $\chi $ denotes a hardening parameter, and $Y=$ $%
Y\left( \chi \right) $ denotes the cohesion. Strain increment $d\mathbf{%
\varepsilon }$ of the matrix can be divided (e.g., \cite{Han and Reddy 1999}%
, \cite{Nikolaevskij 1990}) into elastic $\left( d\mathbf{\varepsilon }%
^{e}\right) $ and plastic $\left( d\mathbf{\varepsilon }^{p}\right) $ parts: 
$d\mathbf{\varepsilon }=d\mathbf{\varepsilon }^{e}+d\mathbf{\varepsilon }%
^{p} $. The plastic strains are determined by non-associative flow rule \cite
{Nikolaevskij 1990} that can be written in the form 
\begin{equation}
\mathbf{e}^{p}=\left[ \mathbf{\sigma }^{f}+\frac{2}{3}\Lambda Y\mathbf{I-}%
\left( 1+\frac{2}{3}\Lambda \alpha \right) \sigma ^{f}\mathbf{I}\right] \dot{%
\lambda},  \label{S210}
\end{equation}
where $\Lambda =\Lambda \left( \chi \right) $ is the dilatancy rate, $%
\mathbf{e}^{p}\equiv d\mathbf{\varepsilon }^{p}\diagup dt$, $\dot{\lambda}%
\equiv d\lambda \diagup dt$. The scalar function $\dot{\lambda}=0$ if $%
C_{\sigma }<0$. Following Sedov \cite[V. 4, pp. 145-147]{Sedov 1971}, in
view of the first and second law of thermodynamics, as applied to the solid
matrix of granular media, we obtain 
\begin{equation}
\left( 1-\phi \right) \rho _{s}T\frac{d_{i}S}{dt}=-\frac{\mathbf{\ q\cdot }%
\nabla T}{T}+\mathbf{\theta :e}^{p}\geq 0,  \label{S360}
\end{equation}
where $\mathbf{\theta =\sigma }^{f}-\left( 1-\phi \right) \rho _{s}\partial
F\diagup \partial \mathbf{\varepsilon }^{p}$, $S$ denotes the specific
entropy, $d_{i}S$ denotes the specific entropy variation due to irreversible
processes, $T$ denotes the absolute temperature of the solid phase, $F\equiv
U-TS=F\left( \mathbf{\varepsilon }^{e},\mathbf{\varepsilon }^{p},T\right) $
denotes the specific free energy, $U$ denotes the internal energy, $\mathbf{q%
}$ is the heat flux. Thus, the energy dissipation is determined by
thermodynamic currents $\mathbf{q}$, $\mathbf{e}^{p}$ and conjugated forces,
according to the bilinear form of (\ref{S360}). In view of Curie principle 
\cite{Gyarmati 1970}, the value of energy dissipation associated with
plastic deformation $D^{p}\equiv $ $\mathbf{\theta :e}^{p}\geq 0$. Hence,
the non-associative flow rule (\ref{S210}) is bound to be equivalent to the
following constitutive equation \cite{Gyarmati 1970}: $\mathbf{e}^{p}\mathbf{%
=L:\theta }$, where $\mathbf{L}$ is a fourth-rank tensor. It is possible on
condition that 
\begin{equation}
\mathbf{\theta =}\left\{ 
\begin{array}{cc}
\mathbf{\sigma }^{f} & if\ \Lambda =0 \\ 
\mathbf{\sigma }^{f}-\frac{Y}{\alpha }\mathbf{I} & if\ \Lambda \neq 0
\end{array}
\right. .  \label{S365}
\end{equation}
Thus, in the case of granular media, the specific free energy $F$ is a
function of the first invariant of plastic strain tensor $\mathbf{%
\varepsilon }^{p}$.

Following Sadovskii \cite[Section 1.2]{Sadovskii 1997}, we assume that there
exists a convex, possibly non-differentiable, function $B\left( \mathbf{e}%
^{p}\right) $ such that $\mathbf{\theta }\in \partial B\left( \mathbf{e}%
^{p}\right) $, where $\partial B\left( \mathbf{e}^{p}\right) $ denotes the
sub-differential (\cite{Panagiotopoulos 1985}, \cite{Sadovskii 1997}) of the
function $B$. It is assumed \cite[Section 1.2]{Sadovskii 1997} that $B\left( 
\mathbf{e}^{p}\right) $ is a positive-homogeneous function of $\mathbf{e}%
^{p} $, i.e., $B\left( b\mathbf{e}^{p}\right) =bB\left( \mathbf{e}%
^{p}\right) $ for $b\geq 0$. The above sub-differential relationship is
equivalent (\cite{Panagiotopoulos 1985}, \cite{Sadovskii 1997}) to 
\begin{equation}
\mathbf{\theta :e}^{p}-B=\underset{\mathbf{e}_{\ast }^{p}}{\max }\left( 
\mathbf{\theta :e}_{\ast }^{p}-B_{\ast }\right) .  \label{S370}
\end{equation}
The right-hand side of (\ref{S370}) is the Young's transformation \cite
{Sadovskii 1997} of $B$, and, in view of the positive-homogeneousness, is
the characteristic function of the convex set $\Upsilon =$ $\left\{ \mathbf{%
\theta }\mid C_{\sigma }\leq 0\right\} $, i.e. the right-hand side of (\ref
{S370}) is equal to zero for $\mathbf{\theta \in \Upsilon }$, and hence $B=$ 
$\mathbf{\theta :e}^{p}$, i.e. $B\equiv D^{p}$. Using the Young's
transformation for the characteristic function of the convex set $\Upsilon $%
, we obtain 
\begin{equation}
D^{p}=\underset{\mathbf{\theta }_{\ast }\in \Upsilon }{\max }\left( \mathbf{%
\ \theta }_{\ast }:\mathbf{e}^{p}\right) .  \label{S380}
\end{equation}
In view of (\ref{S380}) we obtain the desired variational principle: 
\begin{equation}
\left( \mathbf{\theta -\theta }_{\ast }\right) \mathbf{:}d\mathbf{\
\varepsilon }^{p}\geq 0,\quad \mathbf{\theta },\mathbf{\theta }_{\ast }\in
\Upsilon ,  \label{S460}
\end{equation}
where $\mathbf{\theta }$\ and $\mathbf{\theta }_{\ast }$ denote the tensors
associated with actual and, respectively, virtual energy dissipation.

\section{Concluding remarcs}

To fulfill the basic laws of conservation by the mean quantities, the
averages were introduced by different methods. In particular, the velocities
and body forces are mass-averaged, whereas the densities and stresses are
treated as volume averaged. Owing to this approach we succeeded in modeling
of the Reynolds stress tensors under variable densities. Thus, the developed
macroscopic balance equations are applicable for modeling flows through a
permeable medium under a wide range of velocities.

The novel variational principle (\ref{S460}), developed for plastic
deformation of granular media, can be interpreted as the maximum plastic
energy dissipation principle. If a granular medium is idealized to be
cohesionless ($Y=0$), then in view of (\ref{S365}) $\mathbf{\theta =\sigma }%
^{f}$, $\mathbf{\theta }_{\ast }=\mathbf{\sigma }_{\ast }^{f}$, and hence
the novel variational principle (\ref{S460}) coincides with the maximum
plastic work principle \cite{Han and Reddy 1999}.

\section{Appendix\label{Appendix}}

Let us estimate the value of $\omega _{f}$ in (\ref{S75-10}) by considering
the motion of an incompressible viscous fluid through a medium consisting of
straight cylindrical tubes. Let us note that because of incompressibility we
have $\widetilde{e_{f}}$ $\equiv \ \overline{e_{f}}$, $\forall $ $e_{f}$.

First we consider Hagen-Poiseuille flow (see, e.g., \cite[p. 117]
{Schlichting and Gersten 2000}), i.e., a steady-state laminar flow regime.
In polar coordinates the velocity distribution over a cross-section of $i-th$
tube can be written in the form

\begin{equation}
\mathbf{U}_{i}=\frac{\zeta _{i}R_{i}^{2}}{4\mu }\left( 1-\frac{r_{i}^{2}}{%
R_{i}^{2}}\right) \mathbf{k}_{i}\quad 0\leq r_{i}\leq R_{i},  \label{S76-10}
\end{equation}
where $R_{i}$ denotes the radius of $i-th$ tube, $\zeta _{i}$ denotes the
pressure drop, and $\mathbf{k}_{i}$ $\equiv $ $\left( k_{x,i}\right. $, $%
k_{y,i}$, $\left. k_{z,i}\right) $ denotes the unit vector parallel to the
axis of the tube $i$. By virtue of (\ref{S76-10}) we obtain that 
\begin{equation}
\mathbf{J}_{i}\equiv 2\pi \int\limits_{0}^{R_{i}}\mathbf{U}_{i}r_{i}dr_{i}=%
\frac{\pi \zeta _{i}R_{i}^{4}}{24\mu }\mathbf{k}_{i},  \label{S76-12}
\end{equation}
Then clearly 
\begin{equation}
\overline{\mathbf{U}}=\frac{\sum \mathbf{J}_{i}}{\sum \pi R_{i}^{2}}=\frac{%
\sum \zeta _{i}R_{i}^{4}}{8\mu \sum R_{i}^{2}}\mathbf{k}_{i}.  \label{S76-14}
\end{equation}
Let $\xi _{i}=\zeta _{i}R_{i}^{4}$. Then, in view of (\ref{S76-14}), we find 
\begin{equation}
\left( \overline{\mathbf{U}}\right) ^{2}=\frac{\left( \sum \xi
_{i}k_{x,i}\right) ^{2}+\left( \sum \xi _{i}k_{y,i}\right) ^{2}+\left( \sum
\xi _{i}k_{z,i}\right) ^{2}}{64\mu ^{2}\left( \sum R_{i}^{2}\right) ^{2}}.
\label{S76-15}
\end{equation}
By virtue of the elementary inequality, $a^{2}$ $+\ b^{2}$ $\geq $ $2ab$, it
is an easy matter to prove that the numerator in the right-hand side of (\ref
{S76-15}) is bounded above by $\left( \sum \xi _{i}\right) ^{2}$. Hence we
get 
\begin{equation}
\left( \overline{\mathbf{U}}\right) ^{2}\leq \frac{\left( \sum \zeta
_{i}R_{i}^{4}\right) ^{2}}{64\mu ^{2}\left( \sum R_{i}^{2}\right) ^{2}}.
\label{S76-18}
\end{equation}
In a similar manner, we obtain

\begin{equation}
J_{i}^{\ast }\equiv 2\pi \int\limits_{0}^{R_{i}}\left( \mathbf{U}_{i}\right)
^{2}r_{i}dr_{i}=\frac{\pi \zeta _{i}^{2}R_{i}^{6}}{48\mu ^{2}}.
\label{S76-20}
\end{equation}
Then obviously 
\begin{equation}
\overline{\left( \mathbf{U}\right) ^{2}}=\frac{\sum J_{i}^{\ast }}{\sum \pi
R_{i}^{2}}=\frac{\sum \zeta _{i}^{2}R_{i}^{6}}{48\mu ^{2}\sum R_{i}^{2}}.
\label{S76-30}
\end{equation}
In view of (\ref{S76-30}) and (\ref{S76-18}) we obtain from (\ref{S75-10})
that 
\begin{equation}
\omega _{f}\geq \frac{4\sum R_{i}^{2}\sum \zeta _{i}^{2}R_{i}^{6}}{3\left(
\sum \zeta _{i}R_{i}^{4}\right) ^{2}}-1.  \label{S76-50}
\end{equation}
The lower bound of $\omega _{f}$ follows from Cauchy-Schwarz inequality 
\begin{equation}
\left( \sum a_{i}b_{i}\right) ^{2}\leq \sum a_{i}^{2}\sum b_{i}^{2}.
\label{S76-60}
\end{equation}
Setting $a_{i}=R_{i}$ and $b_{i}=\zeta _{i}R_{i}^{3}$, we obtain from (\ref
{S76-50}), in view of (\ref{S76-60}), that $\omega _{f}\geq 1/3$. Notice, $%
\omega _{f}=1/3$ if the tubes are identical in diameter and their axes are
parallel to each other.

Next we consider a turbulent regime. The velocity distribution over a
cross-section of $i$-th tube can be written (e.g., \cite[p. 563]{Schlichting
1968}) in the following form 
\begin{equation}
\mathbf{U}_{i}=U_{m,i}\left( 1-\frac{r_{i}}{R_{i}}\right) ^{1\diagup n}%
\mathbf{k}_{i}\quad 0\leq r_{i}\leq R_{i},\quad n=const,  \label{S76-70}
\end{equation}
where $U_{m,i}$ denotes the maximum velocity. Notice, we assume $n$ $=$ $%
const$ in the empirical formula (\ref{S76-70}), as the exponent $n$ varies
slightly with the Reynolds number $Re$ \cite[p. 563]{Schlichting 1968}. In
the perfect analogy to the deduction of (\ref{S76-18}), (\ref{S76-30}), and (%
\ref{S76-50}), we obtain in the case of turbulent regime that 
\begin{equation}
\left( \overline{\mathbf{U}}\right) ^{2}\leq \frac{4n^{4}\left( \sum
U_{m,i}R_{i}^{2}\right) ^{2}}{\left( n+1\right) ^{2}(2n+1)^{2}\left( \sum
R_{i}^{2}\right) ^{2}},\quad \overline{\left( \mathbf{U}\right) ^{2}}=\frac{%
n^{2}\sum U_{m,i}^{2}R_{i}^{2}}{\left( n+1\right) \left( n+2\right) \sum
R_{i}^{2}},  \label{S76-80}
\end{equation}
\begin{equation}
\omega _{f}\geq \frac{(2n+1)^{2}\left( n+1\right) \left( \sum
R_{i}^{2}\right) \sum U_{m,i}^{2}R_{i}^{2}}{4n^{2}\left( n+2\right) \left(
\sum U_{m,i}R_{i}^{2}\right) ^{2}}-1.  \label{S76-90}
\end{equation}
Let $a_{i}=R_{i}$ and $b_{i}=U_{m,i}R_{i}^{2}$ in (\ref{S76-60}). Then, in
view of (\ref{S76-90}), we obtain

\begin{equation}
\omega _{f}\geq \frac{5n+1}{4n^{2}\left( n+2\right) }.  \label{S76-100}
\end{equation}
In particular, if $n=6$, i.e. $Re=4\cdot 10^{3}$ \cite[p. 563]{Schlichting
1968}, then we find, by virtue of (\ref{S76-100}), that $\omega _{f}\geq
0.027$. Thus, in the case of turbulent regime the minimal value of $\omega
_{f}$\ is far less than in the case of laminar regime.

\end{document}